\documentclass[12pt]{iopart}
\usepackage{graphicx} 
\usepackage{dcolumn} 
\usepackage{bm}
\newcommand{\ignore}[1]{} 
\usepackage{epsfig} 
\usepackage{color}
\usepackage{iopams}
\usepackage{multirow}
  
\begin{document}

\title{Flocking dynamics with voter-like interactions}

\author{Gabriel Baglietto$^1$} 

\author{Federico Vazquez$^1$} 

\address{$^1$ IFLYSIB, Instituto de F\'isica de
  L\'iquidos y Sistemas Biol\'ogicos (UNLP-CONICET), 1900 La Plata,
  Argentina}

\date{\today}

\begin{abstract}
We study the collective motion of a large set of self-propelled
particles subject to  voter-like interactions.  Each particle moves on
a two-dimensional space at a constant speed in a  direction that is
randomly assigned initially.  Then, at every step of the dynamics,
each particle adopts the direction of motion of a randomly chosen
neighboring particle.  We investigate the time evolution of the global
alignment of particles measured by the  order parameter $\varphi$,
until complete order $\varphi=1.0$ is reached (polar consensus).  We
find that $\varphi$ increases as $t^{1/2}$ for short times and
approaches exponentially fast to $1.0$ for long times.  Also, the mean
time to consensus $\tau$ varies non-monotonically with the density of
particles $\rho$, reaching a minimum at some intermediate density
$\rho_{\mbox{\tiny min}}$.  At $\rho_{\mbox{\tiny min}}$, the mean
consensus time scales with the system size $N$ as 
$\tau_{\mbox{\tiny min}} \sim N^{0.765}$, and thus the
consensus is faster than in the case of all-to-all interactions (large $\rho$)
where $\tau=2N$.  We show that the fast consensus, also observed at
intermediate and high densities, is a consequence of the segregation
of the system into clusters of equally-oriented particles which breaks
the balance of transitions between directional states in well
mixed systems.   
\end{abstract}
\maketitle

\section{Introduction}
\label{intro}

Imitation is one of the a fundamental social mechanisms implemented in
most statistical physics models of opinion formation
\cite{Castellano-2009}.  By means of local pairwise interactions
between agents, the imitation mechanism generates an ordering dynamics
that allows to study how a population of individuals reaches
consensus, i e., a state where all individuals share the same opinion.
Among the most studied opinion formation models, the so-called voter
model (VM) is the most basic model that introduces social imitation in
a simple way.  In its original formulation
\cite{Clifford-1973,Holley-1975}, each individual or voter is located
at the site of a $d$-dimensional lattice and can take one of two
possible opinions represented by an up or a down spin ($S=\uparrow$ or
$\downarrow$).  At each time
step of the dynamics, a voter chosen at random simply adopts the
opinion of a randomly chosen nearest neighbor.  This local alignment
dynamics leads to the formation of same-spin domains that growth in
size until one domain takes over the entire lattice and the system
reaches consensus in a time that scales as $N^2$, $N \ln N$ and $N$ in
one, two and three dimensions, respectively
\cite{Krapivsky-1992,Frachebourg-1996}.  This ultimate state is frozen
as spins cannot longer evolve.  The macroscopic dynamic behavior of
the VM can be  understood in terms of its associated Langevin equation
for the magnetization field \cite{Alhammal-2005,Vazquez-2008-b}, whose
Ginzburg-Landau potential is zero.   The VM with an arbitrary number
of opinion states (the multi-state Voter Model) has recently been studied
in \cite{Starnini-2012,Pickering-2016}, where the authors found
interesting properties on the evolution of the number of different
opinions and the consensus times.
The VM has also been applied to the study of other processes
like kinetics of heterogeneous catalysis
\cite{Krapivsky-1992,Frachebourg-1996}, species competition
\cite{Clifford-1973} and ecological diversity
\cite{Durret-1996,Chave-2002,Zillio-2005}.  Several extensions of the
VM have been investigated in the literature, including the presence of
zealots or inhomogeneities 
\cite{Mobilia-2003}, constrained interactions \cite{Vazquez-2003},
non-equivalent states \cite{Castello-2006}, asymmetric transitions or
bias \cite{Antal-2006}, noise \cite{Medeiros-2006} and memory effects
\cite{DallAsta-2007,Stark-2008,Masuda-2010,Woolcock-2017},
among others.   

While all these works assumed that agents are fixed in space, some
recent works have considered the case in which agents are allowed to
move.  In reference \cite{Sousa-2008} Sousa et al. introduced mobility in the
Sznadj Model for opinion formation and found that consensus is always
reached given that agents' diffusion remove the typical frustrations
observed in the static case.  Terranova et al. \cite{Terranova-2014}
studied a multi-state opinion model with moving individuals and showed
that low motility enhances the tendency of the population to adopt
moderate states.  Lipowska and Lipowski \cite{Lipowska-2017} explored
the influence of migrations in language formation by means of a model
called Naming Game in which agents move diffusively on a 
two-dimensional (2D) lattice, finding that the language of agents with
lowest mobility 
is favored and that consensus is slower than in systems without motility.

The models described above assume that the direction of motion
of a given particle is independent on its opinion or language state.
However, in some other social phenomena like the flocking behavior
of large groups of animals such as bird flocks, fish schools or insects swarm 
(see \cite{vicsek2012,marchetti2013,menzel2015} for recent reviews),
the direction of motion of each individual is given by its opinion or
decision.  For instance, Couzin et al. \cite{Couzin-2011} combined
experiments and theory to study collective decision making in fish
schools.  In these experiments, a group of fish is released in a pool
and then each fish decides whether to go to a yellow or a blue target.  Starting
from a minority group trained to have a strong preference for a given
target and that is able to persuade the majority group, they showed that the
addition of a large enough group of uninformed individuals can promote
democratic consensus, returning the control to the majority.  
In one of the seminal works that connected the field of swarming
behavior with that of opinion formation \cite{Huepe-2011} the authors
introduced an non-spatial adaptive network approach to model swarming
experiments with locusts on a ring-shaped arena \cite{Buhl-2006}.
Agents were endowed with one of two possible states (right or left)
representing the two possible moving directions of each locust
(clockwise and anti-clockwise), and updated their states by interacting with
their neighbors in the network.  The model reproduced 
qualitatively the collective properties observed in the experiment,
such as a spontaneous symmetry breaking and a density-driven order-disorder
transition.

In this article we explore the mechanism of social imitation in
flocking dynamics, where the direction of motion of each particle is
associated to its opinion state.  We propose and study a simple
two-dimensional flocking model with voter-like interactions in which,
in a single iteration step, each  
particle adopts the direction of one particle chosen at random within
its interacting neighborhood.  Unlike the models studied in 
\cite{Couzin-2011} and \cite{Huepe-2011} and other related models 
\cite{Martinez-2015}, where the possible moving options 
are binary, all possible directions in the range $(-\pi,\pi]$ are
allowed in our model.  Besides, interactions are not of mean-field
type as in \cite{Huepe-2011}, but rather take into account the 
distance between nearby particles in an euclidean $2D$ space.  If we
think in the example of a flock of birds, the imitation rule can be
interpreted as each bird copying the direction of one close-by bird
even if it is able to see a group of, for instance, $10$ birds around.  
Some recent works have also implemented other types of social
interactions to study flocking, like the majority rule dynamics 
explored in \cite{Kenkre-2007}, where   
interactions are not pairwise as in our model but they include a group 
of neighbors as in the Standard Vicsek Model (SVM)
\cite{Vicsek-1995}.  Also, the model proposed by Chou and Ihle
\cite{Chou-2015} considered the low density limit of the SVM where
particles interact by pairs and both take their average orientations,
unlike in our model where only one of the two particles updates its direction. 
We want to note as well that the flocking voter model (FVM) proposed here
can be framed as a particular type of coevolving network model
\cite{Holme-2006,Vazquez-2008-a,Vazquez-2013}, in which the network of
interactions evolve as particles enter and leave the interaction range
of other particles.  

We show in this article that the  ordering dynamics of the FVM
is much slower than that in the SVM.  Also, the interplay between the
interaction pattern and the alignment dynamics leads to a relaxation
to the final ordered state that largely deviates from the one observed
in mean-field, as well as in lattices with static particles.   As a
consequence, the mean time to reach full polar order (consensus)
exhibits a non-monotonic dependence with the density of particles.
The motion appears to accelerate the polar consensus by a non-trivial
mechanism that breaks the equivalence between particle states
typically observed in the voter dynamics, introducing a net drift from
minority to majority states. 

The article is organized as follows.  We introduce the model and its
dynamics in section \ref{model}.  In section \ref{ordering} we explore
the ordering dynamics. We analyze the consensus times to the final
ordered state in section \ref{consensus}.  Finally, we discuss the results
and state the conclusions in section \ref{conclusions}.

\section{The Model}
\label{model}

We consider a set of $N$ particles that move on a continuous
two-dimensional space $[0,L]^2$ with periodic boundary conditions.
The density of particles $\rho=N/L^2$ is conserved at all times.  At a
given time $t$, the position of particle $i$ is denoted by
$\vec{r_i}^t=(x_i^t,y_i^t)$ and its velocity by $\vec{v_i}^t = (v \cos
\theta_i^t, v \sin \theta_i^t)$, with speed $|\vec{v_i}|=v$ and
direction $\theta_i^t$, for $i=1,..,N$.  That is, all   particles move
at the same speed $v$ but not necessarily in the same direction.  At
each time step of length $\Delta t=1$, each particle $i$ selects a random
neighboring particle $j$ inside a circular region of radius
$R=1$ centered at $\vec{r_i}$, and updates its position and direction
according to 
\numparts
\begin{eqnarray}
\label{update-position}
\vec{r_i}^{t+1} &=& r_i^t + \vec{v_i}^t \Delta t, \\
\label{update-angle}
\theta_i^{t+1} &=& \theta_j^t,
\end{eqnarray}
\endnumparts 

where $\theta_j^t$ is the direction of the particle $j$ at time $t$.
In case particle $i$ has no  neighbors inside the interaction range
$R$, then its direction is not changed.  At $t=0$, positions of
particles are assigned randomly with a uniform distribution inside the
box $[0,L]^2$, while their directions are randomly chosen from the
interval $(-\pi,\pi]$.  Then, each particle moves at a constant speed 
following a given straight path and can update its direction at 
integer times $t=1,2,3,...$, by adopting the direction of a neighbor
chosen at random.  We mention that we are choosing in our model the updating 
rule used in the original version of the Vicsek model 
\cite{Vicsek-1995}, which is known under the name of {\it backward 
update}.  In the backward update the velocity at time $t$ is used 
to obtain the position of a particle at the next time $t+1$ 
(Eq.~\ref{update-position}), whereas in the {\it forward update} 
that position is obtained using the velocity at time $t+1$.  This
ambiguity in the selection of the position  update and their
consequences in the dynamics of the Vicsek model were discussed in 
some works (see for instance \cite{bagliettoBU}).  However, we expect
the qualitative behavior of the FVM to be the same under both updates
at low speeds.  Qualitative differences may appear at high speeds
where the difference between moving a particle before or after
adopting a direction could be very large in some cases.

\section{Ordering dynamics} 
\label{ordering}

Voter-like interactions in the dynamics of the model tend to align the
direction of neighboring particles.  This leads to a local order in the
short run, and to global or macroscopic 
polar order in the long run, as it happens in flocking models with
ferromagnetic interactions like the SVM.  
The ordering properties in these systems are characterized by the 
parameter $\varphi$, defined as  
\begin{equation}
\varphi \equiv \frac{1}{v \, N} \left| \sum_{i=1}^N \vec{v_i} \right|,
\label{phi}
\end{equation}
which is the absolute value of the normalized mean velocity of all
particles.  The order parameter $\varphi$ can vary from $0$ (total
disorder) to $1.0$ (full order).  

\begin{figure}[t]
\vspace{0.5cm}
\centerline{\includegraphics[width=7.5cm]{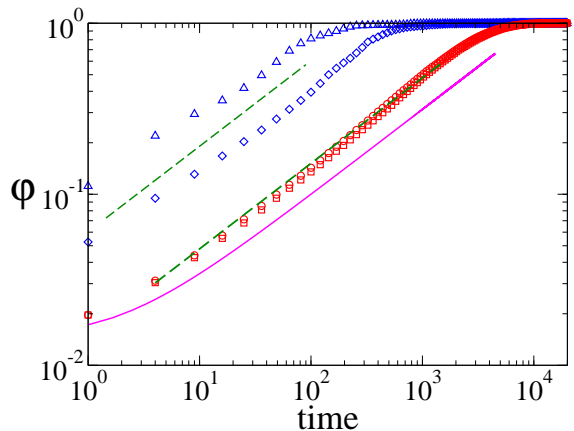}}
\caption{(Color online)  Time evolution of the average order parameter
  $\left< \varphi \right>$ in a system with $N=4000$ particles, with
  speed $v=0.1$.  Blue (upper) symbols correspond to results of the
  SVM at zero noise for particle densities $\rho=21$ (triangles) and
  $\rho=5$ (diamonds).  Red (lower) symbols are results for the FVM
  with densities $\rho=5$ (squares) and $\rho=21$ (circles).  The
  solid line is the theoretical approximation Eq.~(\ref{phivst}),
  while the dashed lines are guides to the eye with slope $1/2$.} 
\label{phi-t}
\end{figure}

In Fig.~\ref{phi-t} we show the temporal behavior of the average value
of the order parameter $\left< \varphi \right>$ for speed $v=0.1$, and
two different densities.  Averages were done over $10^4$ independent
realizations of the dynamics, as in most  plots shown the article
unless stated.  The two lower curves correspond to the FVM.  We also 
show, for comparison, $\left< \varphi \right>$ for the SVM at zero
noise (two upper curves).  We observe that, in all cases, full 
order $\varphi=1.0$ is eventually achieved in the long run.  In this
final ordered state all particles move in the same direction, thus no
more direction updates are possible.  Although this final state is not
frozen because particles continue moving, it is analogous to the
consensus state in the VM.  The ordering in the FVM is characterized
by an initial increase of
$\left< \varphi \right>$ as a power law in time, with an exponent
close to $1/2$ (lower dashed line) and a final exponential approach to
$1.0$.  Even though the approach to complete order in the FVM is much
slower than that in the noiseless SVM, we see that the SVM also exhibits an
initial algebraic increase of $\left< \varphi \right>$ with time, with
an exponent similar to $1/2$ (upper dashed line). 

\begin{figure}[t]
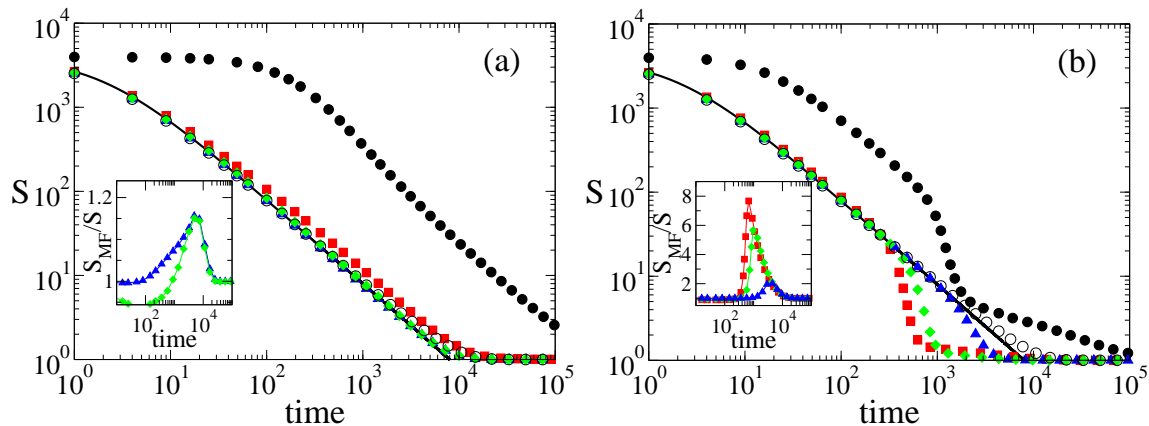

\vspace{0.5cm}
\includegraphics[width=7.5cm]{Fig2a.eps}
\includegraphics[width=7.5cm]{Fig2b.eps}
\caption{Mean number of different directions $S$ vs time for a system of $N=4000$ particles, speeds $v=0.1$ (a)
  and $v=20$ (b), and densities $\rho=0.06$ (circles), $2$ (squares),
  $5$ (diamonds) and $50$ (triangles). Empty circles correspond to
  numerical results of $S$ on a fully-connected system
  ($S_{\mbox{\tiny MF}}$), while solid lines are 
  the analytical expression Eq.~(\ref{Svst}).  Insets: Plots of
  $\frac{S_{\mbox{\tiny MF}}(t)}{S(t)}$ as function of time for
  densities $\rho = 5$ and $50$ in panel (a), and $\rho = 2$, $5$ and
  $50$ in panel (b).  The peaks of
  the curves make evident the departure of the FVM dynamics from the
  mean field case.}  
\label{S-t}
\end{figure}

As we see below, the behavior of $\varphi$ is closely related to that
of the mean number of different directions $S(t)$ at time $t$ shown in
Fig.~\ref{S-t}.  Initially, all directions are different as they are
randomly assigned, and thus $S(t=0)=N$.  Then, for densities $\rho \ge
2$ we see that $S$ decreases very slowly during an initial transient
of order $N$, as 
\begin{equation} 
S(t) \simeq \frac{N}{1+t/2},
\label{Svst}
\end{equation}
represented by a solid line in Fig.~\ref{S-t}.  This result was
derived analytically in \cite{Starnini-2012} following a Master
Equation approach and in \cite{Pickering-2016} using generating
functions, for the  
multi-state voter model under sequential update on a complete graph
(all-to-all interactions).  In 
the synchronous version of the model that we use here time is
rescaled by a factor $1/2$.  In the final regime $S$ relaxes
exponentially fast to $1.0$, corresponding to the single direction of
full order (see Fig.~\ref{S-t}).  
We note that the decrease of $S$ is monotonic at all times.  This is
because some directions may not be copied by any particle in a single
step (mainly those directions followed by few particles), and thus these
directions disappear from the system.  Then, given that new directions are never
created in the voter dynamics, $S$ just decreases monotonically with
time.   

Solid lines in Fig.~\ref{S-t} correspond to Eq.~(\ref{Svst}), which
reproduces quite well the evolution of $S$ from simulations on a
fully-connected system (all-to-all interactions), represented by empty
circles.  We also see that these mean-field (MF) curves agree very
well with simulations on 2D for large particle densities.  Indeed, as
the box's length $L=\sqrt{N/\rho}$ decreases with $\rho$, the MF limit
of all-to-all interactions is achieved as $\rho \to \infty$,
independent on $v$, because $L$ approaches the interaction range
$R=1$.  Deviations from  MF are evident for very low values of $\rho$
(see $\rho=0.06$ curve), where the number of neighbors is very small
and local interactions rule the dynamics.  We also observe that for
low speed $v=0.1$ [Fig.~\ref{S-t}(a)] the agreement between MF and 2D
is quite good for $\rho \ge 2$, but for high speeds $v=20$
[Fig.~\ref{S-t}(b)] 2D results depart from the MF curve at a time
around $10^3$.  We shall discuss this fast decay of $S$ for large
speeds in section \ref{consensus}.         

\begin{figure}[t]
\vspace{0.5cm} \centerline{\includegraphics[width=7.5cm]{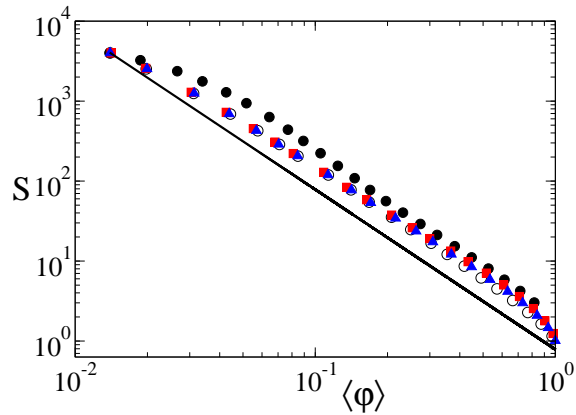}}
\caption{Average number of different directions $S$ vs average order
  parameter $\langle \varphi \rangle$ for a system of $N=4000$
  particles, speed $v=0.1$, and densities $\rho=0.06$ (filled
  circles), $5$ (squares) and $50$ (triangles).  Empty circles
  correspond to numerical results on a fully-connected system, while
  the solid line  is the analytical estimation from
  Eq.~(\ref{phivsS}).}
\label{phi-S}
\end{figure}

An approximate expression for the relation between $\varphi$ and $S$
can be obtained by assuming that the directions of particles are
randomly distributed at all times, as we describe below.  Even though
this assumption is only strictly valid at $t=0$ we shall see that it
also works reasonable well for longer times.  Let's consider that, at
a given time $t$, there are $S(t)$ different independent
directions in the system of particles, drawn uniformly in the
$(-\pi,\pi]$ range.  The number of different directions at time $t$
  may vary between realizations, but here we assume that they are all
  equal to its mean number  $S(t)$.  Then, we can rewrite the order
  parameter at time $t$ from Eq.~(\ref{phi}) in the form
\begin{equation}
\varphi(t) = \frac{1}{v \, N} \left| \sum_{k=1}^{S(t)} n_k(t) \,
\vec{v_k} \right|,
\label{phi-2}
\end{equation}
where the sum is over the different particles' directions $\theta_k^t$
labeled by the index $k$, which runs from $1$ to its total number
$S(t)$.  The number of particles moving in the direction $\theta_k^t$
is denoted by $n_k(t)$ and is normalized at all times $\left(
\sum_{k=1}^{S(t)} n_k(t) =N \right)$.   As particles are initially
assigned random directions, we have that $S(0)=N$ and $n_k(0)=1$ for
all $k=1,..,N$.  The evolution of the occupation numbers $n_k$ is not
trivial, thus we make an approximation and assume that at all times $t
> 0$ they are all equal to its mean value  $n_k(t) \simeq N/S(t)$
($k=1,..,S$).  Therefore, Eq.~(\ref{phi-2}) can be written as 
\begin{equation}
\varphi(t) \simeq \frac{1}{v \, S(t)} \left| \sum_{k=1}^{S(t)}
\vec{v_k} \right| = | \vec{V} |,
\label{phi-3}
\end{equation}
where we have defined the resulting vector 
$\vec{V} \equiv \frac{1}{v \, S(t)} \sum_{k=1}^{S(t)} \vec{v_k}$.
Given the assumption that all velocities $\vec{v_k}$ are independent
and uniformly distributed in $(-\pi,\pi]$, the respective $x$ and $y$ 
components of $\vec{V}$, $V_x$ and $V_y$, are uncorrelated.  Then,
from the central limit theorem we know that, in the $S \gg 1$ limit,
$V_x$ and $V_y$ are normally distributed with zero mean 
($\langle V_x \rangle=\langle V_y \rangle=0$)
\begin{equation}
P(V_x) = \frac{1}{\sqrt{2 \pi} \sigma_x} e^{-\frac{V_x^2}{2 \sigma_x^2}}, 
~~~~
P(V_y) = \frac{1}{\sqrt{2 \pi} \sigma_y} e^{-\frac{V_y^2}{2 \sigma_y^2}}, 
\end{equation}
and equal variance $\sigma_x^2=\sigma_y^2=\sigma^2=1/(2 S)$, and thus 
$|\vec{V}|$ is characterized by the Rayleigh distribution 
\begin{equation}
P(| \vec{V} |) = \frac{| \vec{V} |}{2 \pi \sigma^2} 
e^{-\frac{| \vec{V} |^2}{2 \sigma^2}} .
\end{equation}
Now, we can calculate the average value of the order parameter by
implementing polar coordinates as
\begin{equation}
\langle \varphi(t) \rangle = \langle | \vec{V} | \rangle = 
\int_{-\pi}^{\pi} d \theta \int_{0}^{\infty} d | \vec{V} | ~ 
\frac{| \vec{V} |^2}{2 \pi \sigma^2} e^{-\frac{| \vec{V} |^2}{2
    \sigma^2}} = \sqrt{\frac{\pi}{2}} \, \sigma.
\end{equation}
Finally, using $\sigma=1/\sqrt{2S}$ we arrive to
\begin{equation} 
\left<\varphi\right> \simeq \frac{\sqrt{\pi}}{2} S^{-1/2}.
\label{phivsS}
\end{equation}
In Fig.~\ref{phi-S} we plot $\langle \varphi \rangle$ vs $S$ obtained
from computational simulations on a fully-connected system (empty
circles) and on a 2D system for $v=0.1$ and different densities, and
compare with the behavior predicted by Eq.~(\ref{phivsS}) (solid
line).  At $t=0$ ($S=N$) the expression from Eq.~(\ref{phivsS}) works
very well because all velocities are actually randomly distributed,
and thus the occupation number distribution is uniform ($n_k=1$ for
all $k=1,..,N$).  Then, as groups of particles start to have the same
direction the $n_k$ distribution deviates from uniform, and thus the
assumption  $n_k(t) \simeq N/S(t)$ for all $k$ implemented above does
not hold any more. However, for short times --or large values of $S$--
the uniform approximation still works quite well, and the relation
between $\varphi$ and $S$ for 2D systems is well described by
Eq.~(\ref{phivsS}), as we can see in Fig.~\ref{phi-S}.    

Combining Eqs.~(\ref{Svst}) and (\ref{phivsS}) we arrive to the
following approximate expression for the time dependence of $\langle
\varphi \rangle$
\begin{equation}
\left< \varphi \right> \simeq \frac{\sqrt{\pi}}{2} \left(
\frac{1+t/2}{N} \right)^{1/2},
\label{phivst}
\end{equation}
indicated by a solid line in Fig.~\ref{phi-t}.  Even though there are
discrepancies between the numerics and the analytical expression
Eq.~(\ref{phivst}) for $t > 1$, the algebraic increase $\left< \varphi
\right> \sim \sqrt{t}/N$ for intermediate times (dashed line)
predicted by Eq.~(\ref{phivst}) seems to hold quite well for both
particle densities. In Fig.~\ref{phi-S} we can see that for a fixed
value of $S$, 2D simulations (filled symbols) show more orientational
order than the corresponding MF simulation (empty circles), as
$\langle \varphi \rangle$ is larger in the former case.  This is a
consequence of the fact that in 2D the different directions of
particles at a given time are not uniformly distributed in the
$(-\pi,\pi]$ interval.  Instead, longer lasting directions tend to be
  similar.  This happens because orthogonal directions quickly tend to
  annihilate each other, given that interactions (and alignments) are
  more frequent between particles with high relative cross section.
  This can be seen in the $time=6800$ frame of Fig.~\ref{clus} where
  the two largest clusters move almost parallel to each other.

\section{Consensus times}
\label{consensus}

\begin{figure}[t]
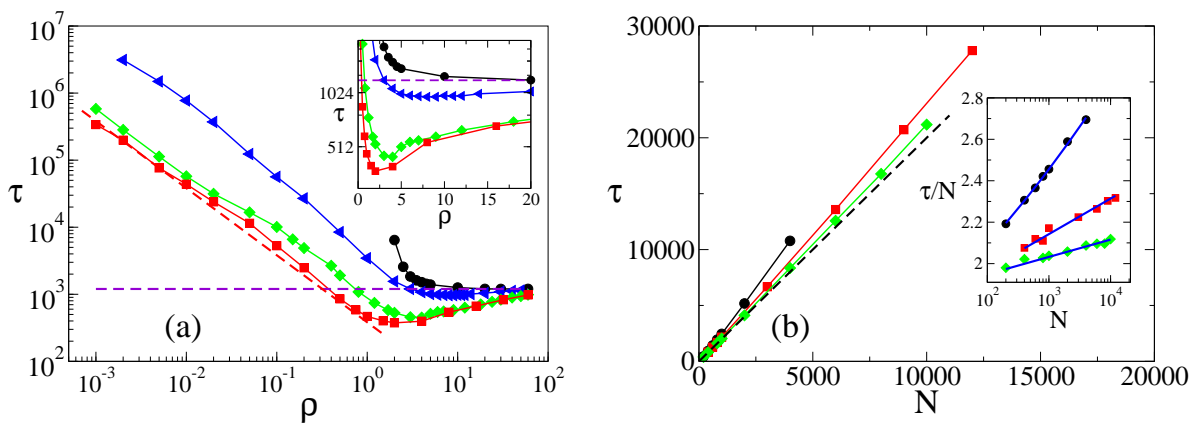

\begin{center}
\vspace{0.5cm}
\begin{tabular}{cc}
\hspace{-1.0cm}
\includegraphics[width=5.0cm, bb=70 -20 550 550]{Fig4a.eps}
& \hspace{2.5cm} 
\includegraphics[width=5.0cm, bb=70 -20 550 550]{Fig4b.eps} 
\end{tabular}
\caption{(a) Mean consensus time $\tau$ vs particle density $\rho$
  for a system of $N=600$ particles and speeds 
  $v=0$ (circles), $v=0.1$ (triangles), $v=5$ (diamonds) and $v=20$
  (squares).  The horizontal dashed line denotes the mean field
  consensus time $\tau_{\mbox{\tiny MF}} \simeq 2N=1200$, while the upper dashed
  line is the approximation from Eq.~(\ref{tau-N-rho}).
  Inset: Same data on a linear-log scale showing the 
  non-monotonic behavior regime.  (b) $\tau$ vs $N$ for $v=0$ and
  densities $\rho=5$ (circles),  $\rho=10$ (squares) and $\rho=21$
  (diamonds).  The dashed line is the expression $\tau = 2N$.  Inset:
  $\tau/N$ vs $N$ on a log-linear scale.  Solid lines are the best
  linear fits $A(\rho) + B(\rho) \ln N$, with coefficients $A = 1.285,
  1.668$ and $1.817$ and slopes $B = 0.170, 0.069$ and $0.032$ for
  densities $\rho = 5, 10$ and $21$, respectively.}
  \label{tau}
\end{center}
\end{figure}

A magnitude of interest in models that exhibit a complete order is the
time to reach the final ordered state or consensus time.  In
Fig.~\ref{tau}(a) we plot the mean consensus time $\tau$ over many
realizations as a function of the particle density $\rho$, for a
system of $N=600$ particles and different speeds.  We need to mention
that there was a very small fraction of realizations that did not reach
consensus, specially at low densities.  Those realizations were not
considered in the calculation of $\tau$.  In the static case
scenario $v=0$ (circles) $\tau$ decreases with $\rho$ and approaches
the MF value ($\tau_{\mbox{\tiny MF}} \simeq 2N=1200$) for large $N$
\cite{Starnini-2012,Pickering-2016} (horizontal dashed line).  As
discussed in section  \ref{ordering}, the MF limiting case is obtained
when $L \le \sqrt{2}$  and thus each particle falls in the interaction
range of any other  particle.  Therefore, for $N=600$ this happens
when $\rho$ overcomes the value $N/2=300$.  However, already for $\rho
\simeq 50$ is $\tau \simeq \tau_{\mbox{\tiny MF}}$ and the system
behaves as in MF. In Fig.~\ref{tau}(b) we show the dependence of
$\tau$ with the number of particles for $v=0$ and different
densities.  For high densities, $\tau$ approaches the MF linear
behavior $\tau_{\mbox{\tiny MF}} \simeq 2N$ (dashed line), but for low
densities there are logarithmic deviations consistent with the 2D
behavior $\tau \sim N \ln N$ \cite{Krapivsky-1992,Frachebourg-1996}.
This crossover between 2D and MF can be better seen in the inset of
Fig.~\ref{tau}(b), where we show $\tau/N$ vs $N$ on a log-linear scale
to capture logarithmic corrections.  We observe that data points fall
on a straight line with density dependent $y$-coordinate $A(\rho)$ and
slope $B(\rho)$.  As $\rho$ increases, $B$ goes to zero and $A$
approaches $2.0$, recovering the MF behavior.  

\begin{figure}[t]
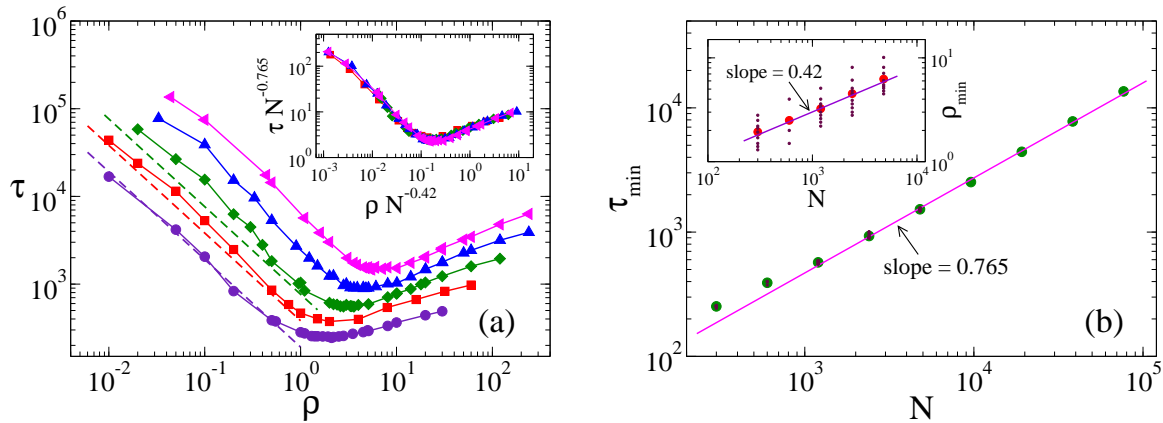

\begin{center}
\vspace{0.5cm}
\begin{tabular}{cc}
\hspace{-1.0cm}
\includegraphics[width=5.0cm, bb=70 -20 550 550]{Fig5a.eps}
& \hspace{2.5cm} 
\includegraphics[width=5.0cm, bb=70 -20 550 550]{Fig5b.eps} 
\end{tabular}
\caption{(a) $\tau$ vs $\rho$ for speed $v=20$ and system sizes
  $N=300$ (circles), $600$ (squares), $1200$ (diamonds), $2400$ (up
  triangles) and $4800$ (left triangles).  Dashed lines are the
  analytic approximation $2N/\pi \rho$ at low densities
  [Eq.~(\ref{tau-N-rho})], for $N=1200$, $600$ and $300$ (from top to
  bottom).  Inset: Rescaled $\tau$ vs 
  rescaled $\rho$ showing the collapse of the curves.  (b)  Auxiliary
  plots showing the regression lines with the exponents $\gamma=0.765 \pm
  0.018$ and $\beta=0.42 \pm 0.09$ used in the inset of panel (a).
  Dots in the main plot and the inset correspond to the set of points 
  $(\rho_{\mbox{\tiny min}},\tau_{\mbox{\tiny min}})$ around the
  minima of the $\tau$ vs $\rho$ curves of panel (a), while circles
  represent respective average values for each $N$.  Main: Minimum
  mean consensus time $\tau_{\mbox{\tiny min}}$ vs $N$.  The straight
  line is the power-law fitting function 
  $\tau_{\mbox{\tiny min}} = 2.378 \, N^{0.765}$ for the 
  range $2400 \le N \le 76800$.  Inset: Density at the minimum value
  of $\tau$, $\rho_{\mbox{\tiny min}}$, vs $N$.  The straight line is the best fit 
  $\rho_{\mbox{\tiny min}} = 0.16  \, N^{0.42}$ to the data.}
\label{tau-rho}
\end{center}
\end{figure}

The dynamic case scenario $v>0$ is different from the static case
$v=0$, as $\tau$ exhibits a non-monotonic behavior with $\rho$.  
The mean consensus time $\tau$ is much larger than 
$\tau_{\mbox{\tiny MF}}$ for low densities, but it decays to values
smaller than $\tau_{\mbox{\tiny MF}}$ as $\rho$ 
increases and, after reaching a minimum value $\tau_{\mbox{\tiny min}}$, increases
and saturates at the MF value.  As expected, $\tau$ becomes
independent of $v$ at high densities.  The behavior of $\tau$ with the
system size $N$ also shows interesting properties, as we see in
Fig.~\ref{tau-rho}(a) where we plot $\tau$ vs $\rho$ for a fixed speed
$v=20$ and several values of $N$.  The shape of all curves are
non-monotonic, and the location of the minimum $\rho_{\mbox{\tiny min}}$ and its
value $\tau_{\mbox{\tiny min}}$ increase with $N$.  When the data
corresponding to each size $N$ is shifted by the factors $N^{0.42}$
and $N^{0.765}$ on the $x$
and $y$ axis, respectively, all curves collapse into a single 
curve [see inset of Fig.~\ref{tau-rho}(a)].  
To obtain these scaling exponents we first selected a set
of data points $(\rho_{\mbox{\tiny min}},\tau_{\mbox{\tiny min}})$
around the minimum of each curve, over an interval of $\rho$ where the
curve $\tau(\rho)$ is approximately flat.  The sets 
$(\tau_{\mbox{\tiny min}},N)$ and $(\rho_{\mbox{\tiny min}},N)$ which
consisted on $43$ points each are plotted by dots in Fig.~\ref{tau-rho}(b) and
its inset, respectively, while circles correspond to average 
values for each $N$.  By doing a linear regression analysis on a
double logarithmic scale, 
$\ln(\rho_{\mbox{\tiny min}}) = \alpha + \beta \ln(N)$, we 
obtained the estimators $\hat \beta = 0.42$ and $\hat \alpha = -1.83$ that
provide the best fit to the $(\rho_{\mbox{\tiny min}},N)$ data, and
their associated error bars
given by the $95 \%$ confidence intervals $\beta = 0.42 \pm 0.09$ and 
$\alpha = -1.83 \pm 0.64$, respectively.  Then, we used the power-law
best fit $\rho_{\mbox{\tiny min}} \simeq 0.16 \, N^{0.42}$ to estimate 
$\rho_{\mbox{\tiny min}}$ for larger values of $N$ (up to $N=76800$) and 
calculated $\tau_{\mbox{\tiny min}}$ for those estimated minima.  The results are
plotted in Fig.~\ref{tau-rho}(b) where we observe that $\tau_{\mbox{\tiny min}}$
follows an asymptotic power law in the interval $2400 \lesssim N \lesssim
76800$, with an exponent $\gamma = 0.765 \pm 0.018$ 
(error bars were estimated using $25$ data points for $N=2400, 4800,
9600, 19200, 38400$ and $76800$).  
We note that the scaling $\tau_{\mbox{\tiny min}} \sim N^{0.765}$ is
consistent with a consensus that is reached faster than in MF, where
$\tau$ grows linearly with $N$. 

The peculiar shape of $\tau$ vs $\rho$ is a consequence of the non-trivial
interplay between particles' speed and the propagation of directions
over the space.  We understand that the approach to consensus depends
on the relation between the time scales associated to these  two
processes --convection and diffusion--, which vary with $\rho$ and
$v$.  Below we explore in more detail the origin of the non-monotonic
behavior of $\tau$ with $\rho$.  We first provide some scaling arguments that
explain the behavior of $\tau$ in the limit of small densities and
large speeds, and then study the clustering dynamics at intermediate
and high densities.

\subsection{Limit of small densities and high speeds}

Let us consider the case where particles move fast, and so they travel
such a long distance in each time step that their distribution remains
nearly uniform over the box.  This is so because the
backward update \cite{bagliettoBU} used in simulations involves two consecutive
events: each particle first aligns its direction with a neighboring
particle and then moves a distance $v$ with the old direction.  As
a consequence, for large speeds ($v \gtrsim L$) aligned particles
could be at any  
distance from each other, leading to a quite uniform spatial
pattern.  Therefore, we can assume that the system remains well mixed at high
speeds and thus interactions are as in MF.
Then, at each time step, direction updates take place only
among those particles that have at least one neighboring particle
inside their interaction range.  For each particle this happens with
probability $p=1-(1-\pi \rho/N)^{N-1} \simeq \pi \rho$ for 
$\pi \rho \le 1 \ll N$, assuming that particles are uniformly distributed.  The 
fact that particles not always update their directions at every time
step introduces a time delay in the alignment dynamics, which rescales
the MF consensus time $\tau_{\mbox{\tiny MF}} \simeq 2N$ by a factor
of $1/p$ (the mean number of attempts between interactions).
Therefore, we arrive to 
\begin{equation}
\tau \simeq \frac{2N}{\pi \rho} ~~~\mbox{for $\rho \le \frac{1}{\pi}$}. 
\label{tau-N-rho}
\end{equation}
We observe in Fig.~\ref{tau}(a) that the expression from
Eq.~(\ref{tau-N-rho}) (dashed line) gives a reasonable estimation of
$\tau$ for high speed $v=20$ in the limit of low densities.  The
power-law decay $\tau \sim \rho^{-1}$ at low $\rho$ is also
observed for speed $v=5$.  Figure~\ref{tau-rho}(a) shows how
Eq.~(\ref{tau-N-rho}) performs as the system size is varied (dashed
lines).  We see that for densities in the range $0.01 < \rho < 0.5$
the approximation is very good for $N=300$ (bottom line), but it
becomes worst for larger system sizes $N=600$ (middle line) and
$N=1200$ (top line).  This is because by increasing $N$, while keeping
$\rho$ and $v$ fixed, the box's length $L=\sqrt{N/\rho}$ increases
and thus the ratio $v/L$ eventually becomes small.  Therefore, we
expect that for large enough system sizes the well-mixed system
assumption $v \gtrsim L$ does not hold any more, and thus the MF
approximation is not longer valid.  We also speculate that, for
a fixed $N$, the agreement becomes better as 
$\rho$ decreases and gets far from the minimum, as we can see for the
$N=600$ curve of Figs.~\ref{tau}(a) and \ref{tau-rho}(a).  We observe
as well that Eq.~(\ref{tau-N-rho})
underestimates the numerical data at intermediate densities, where the
formation of clusters plays an important role, as we show in the next
subsection.

\subsection{Clustering dynamics at intermediate and high densities}

\begin{figure}[t]
\vspace{0.5cm}
\centerline{\includegraphics[width=7.5cm]{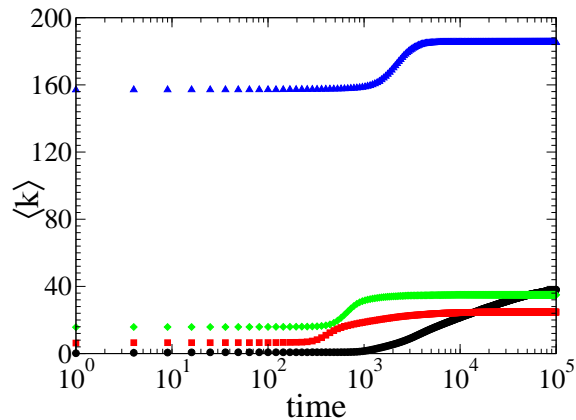}}
\caption{Time evolution of the mean number of interacting neighbors
  $\left< k \right>$, for $N=4000$, speed $v=20$ and densities $\rho=0.06$
  (circles), $\rho=2$ (squares), $\rho=5$ (diamonds) and $\rho=50$
  (triangles).}  
\label{k-t}
\end{figure}

\begin{figure}[t]
\begin{center}
\vspace{0.5cm}
\begin{tabular}{cc}
\hspace{-1.0cm}
\includegraphics[width=4.0cm, bb=70 -20 550 550]{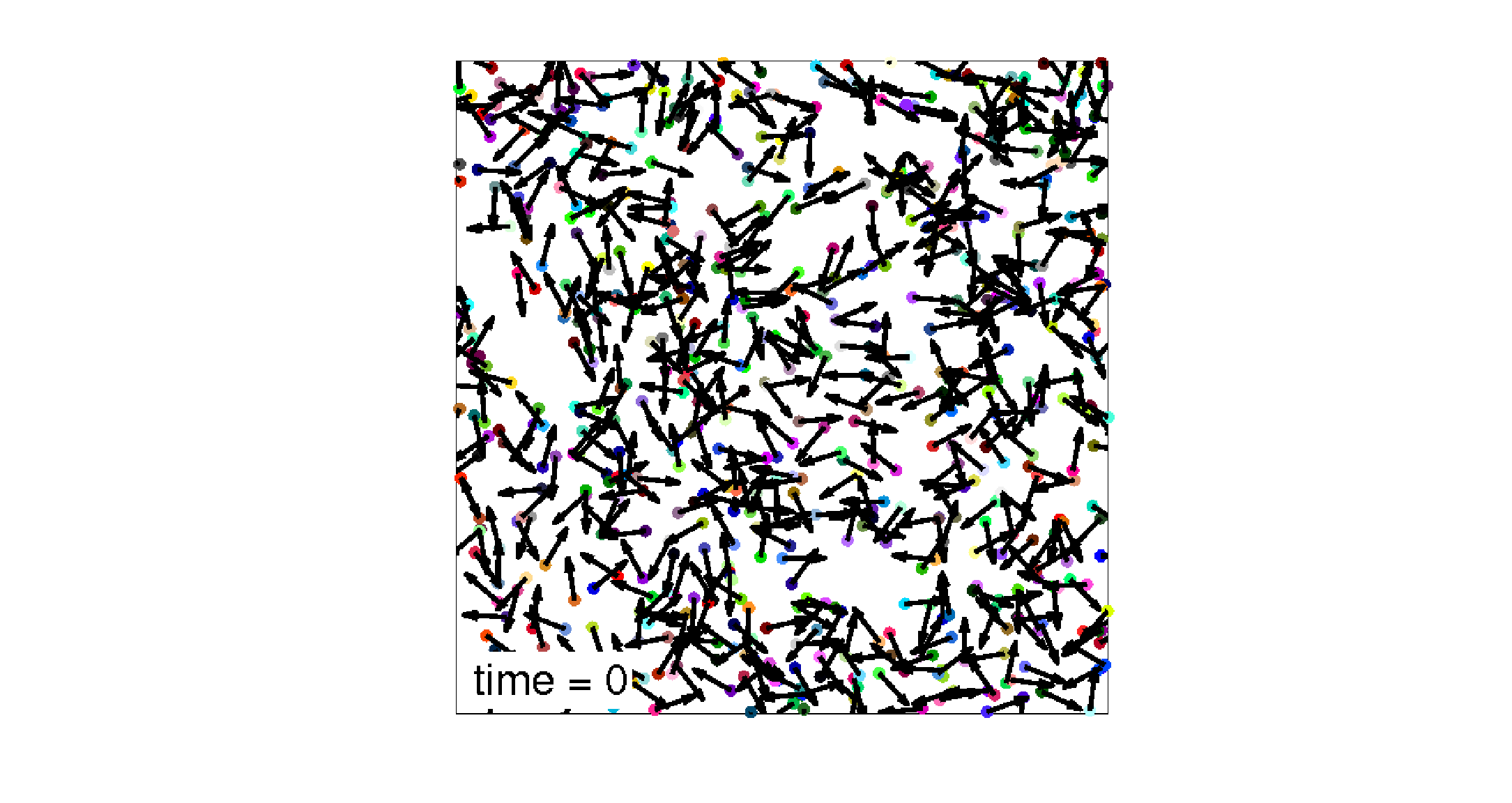}
& \hspace{2.0cm} 
\includegraphics[width=4.0cm, bb=70 -20 550 550]{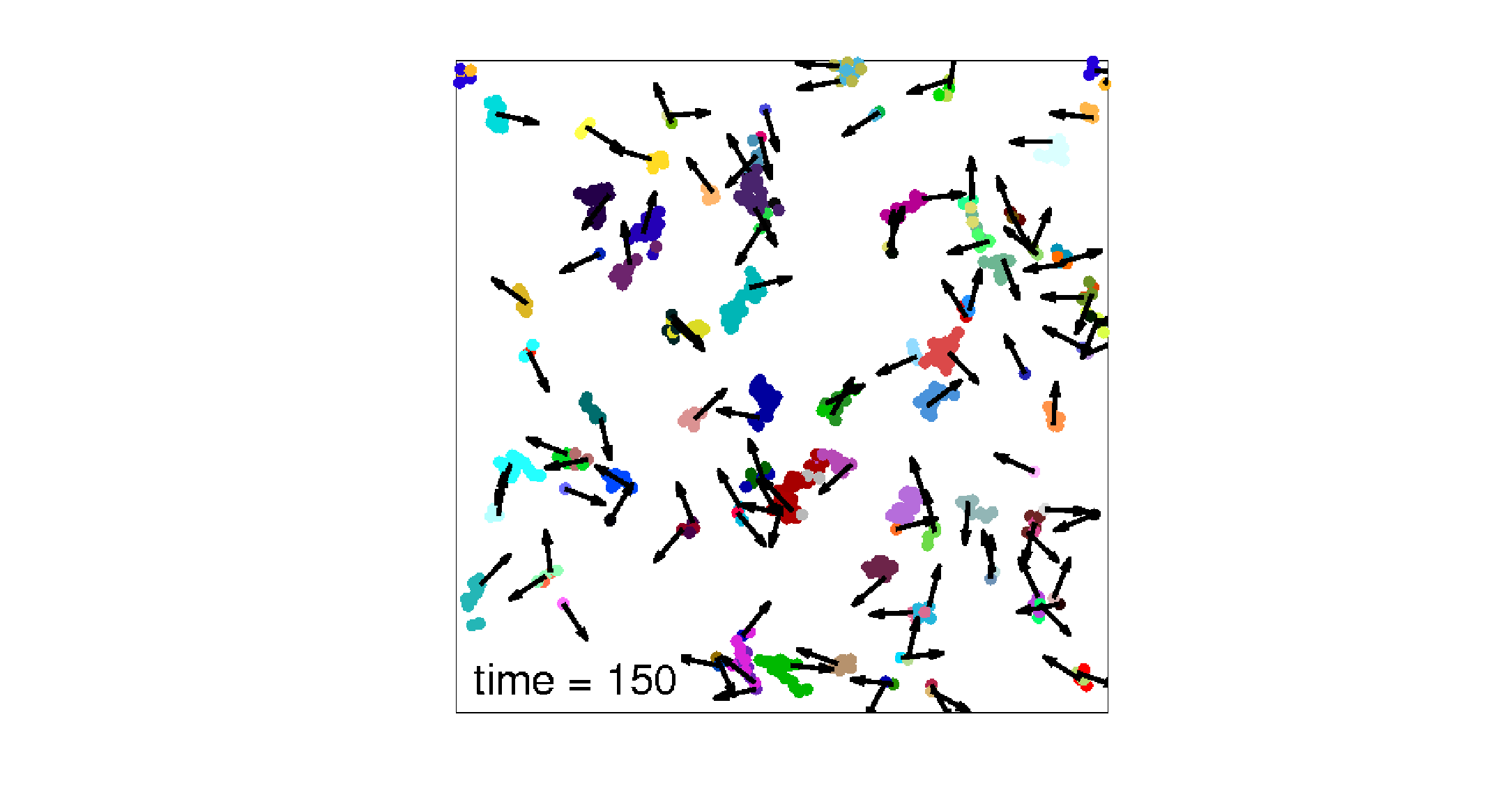} \vspace{0.3cm} \\
\hspace{-1.0cm}
\includegraphics[width=4.0cm, bb=70 0 550 550]{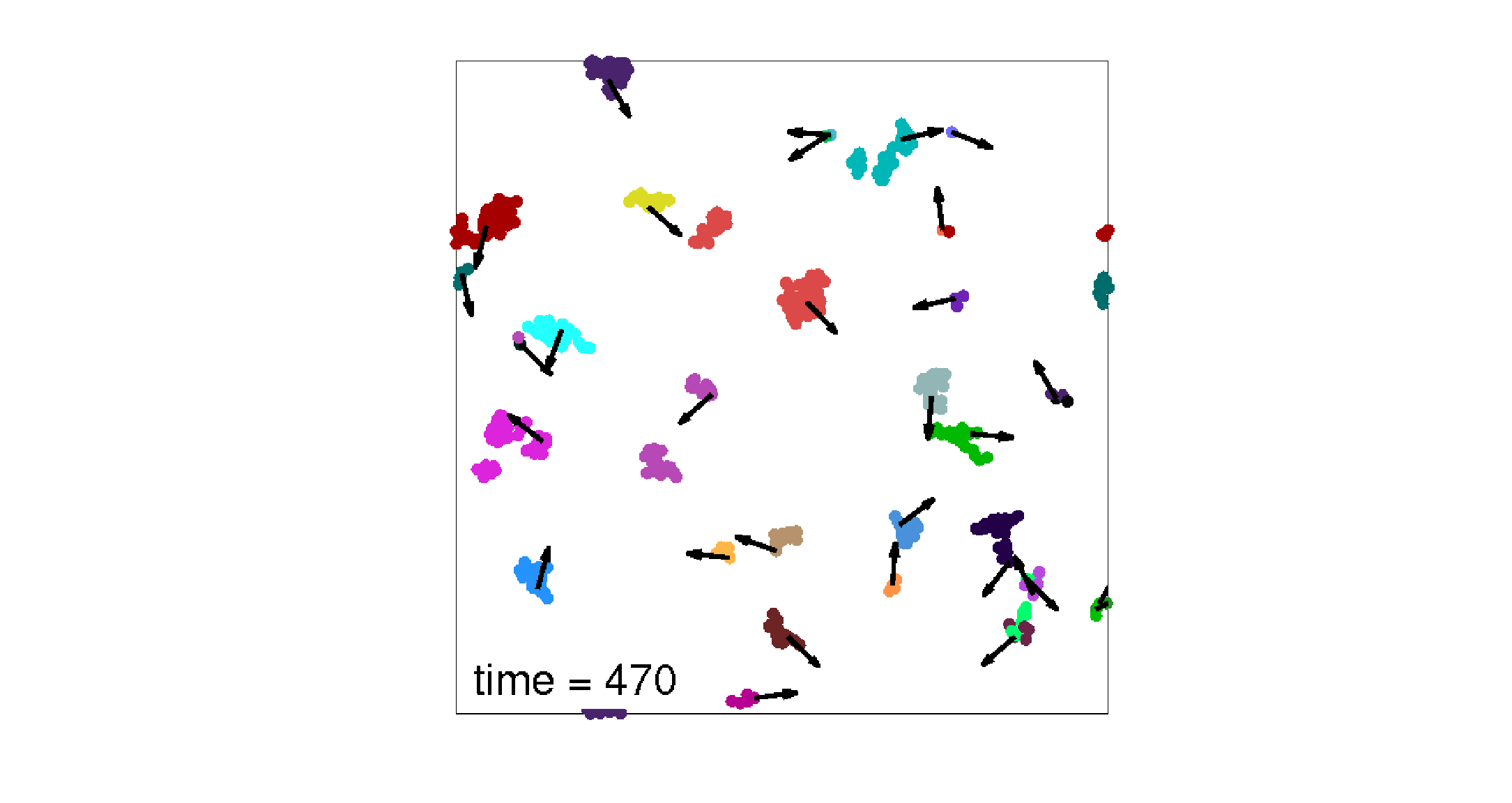}
& \hspace{2.0cm} 
\includegraphics[width=4.0cm, bb=70 0 550 550]{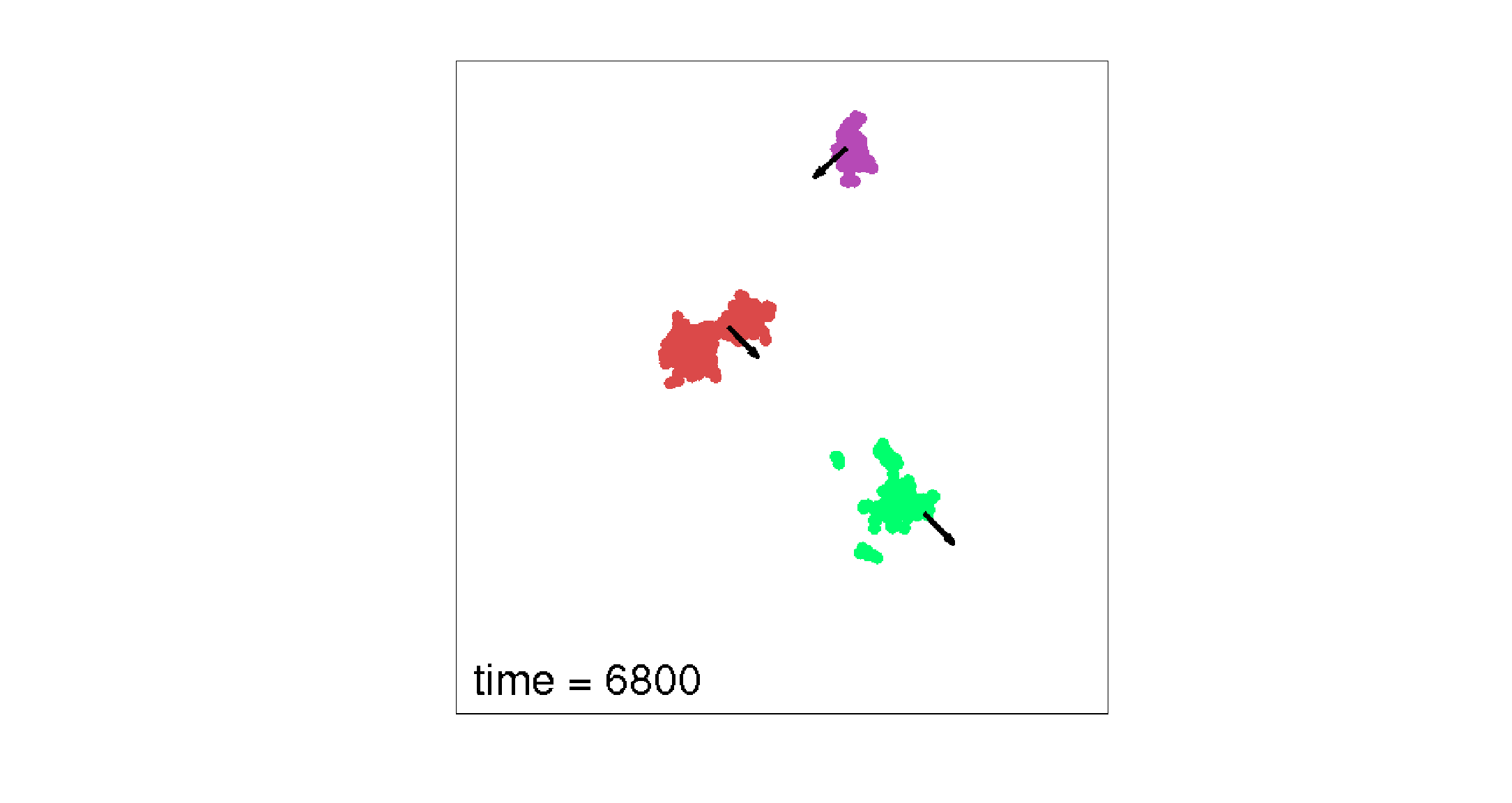} 
\end{tabular}
\caption{(Color online) Snapshots showing the configuration of the 
  system at different times, composed by $N=600$ particles with a
  density $\rho=0.1$ and speed $v=0.1$.
  Particles are depicted by filled circles of radius 
  $1.0$, thus interacting particles overlap in this
  scale.  Arrows indicate the direction of motion of each cluster,
  shown in a particular color.}  
\label{clus}
\end{center}
\end{figure}

For intermediate densities, the consensus turns to be faster than in
MF.  This is consistent with the behavior of the number of different
directions $S$ in Fig.~\ref{S-t}.  There we can see that $S$ decays as
in MF ($S \sim N/t$) during an initial transient, but then it starts
to decay faster (exponentially), leading to a consensus time that is
smaller than $\tau_{\mbox{\tiny MF}}$.  This effect is more pronounced
for large speeds [see Fig.~\ref{S-t}(b)].   As we explain below, the
departure observed in $S$ from a power law to an exponential decay is
caused by a dynamic reordering of the spatial pattern of interactions,
from a well mixed system to structures localized in space.  To explore
this in more detail, we studied the time evolution of the mean number
of neighbors  $\left< k \right>$ (mean degree).  Results are shown in
Fig.~\ref{k-t} for $N=4000$ (the same as in Fig.~\ref{S-t}), $v=20$
and various densities, where we observe that  $\left< k \right>$
starts to increase very slowly from its initial value $\pi \rho$
--corresponding to a uniform distribution of particles-- until it
saturates for long times when the system reaches consensus.  This
increase in the number of neighbors suggests that particles aggregate
into spatial clusters.  Indeed, when two particles are less than a
distance $R=1$ apart they can align their directions and then move
together until one of them changes direction by interacting with a
third particle.  Thus, one can see that alignment implies the
``sticking" of nearby particles, forming large sets of particles that
move together in the same direction.  These structures can be seen in
Fig.~\ref{clus} for a system of $N=600$ particles, with $v=0.1$ and
$\rho=0.1$.  The panels represent snapshots of the system at different
times, showing the transition from a uniform distribution of particles
moving in random directions at $t=0$ (a), to a spatial segregation
into clusters of particles with the same direction at $t=150$ (b) and
$t=470$ (c), and to a quasi-consensus state at $t=6800$.  The
segregation occurs for higher densities as well, but clusters may
overlap when densities are very high.  

By comparing Fig.~\ref{S-t}(b) with Fig.~\ref{k-t}, we observe that
the deviation of $S$ from the MF value starts  approximately when
$\langle k \rangle$ starts to be significantly larger than its initial
value $\pi \rho$; for instance at time $t \simeq 400$ for $\rho=2$.
This indicates that the formation of clusters speeds up the
dynamics and the approach to consensus.  It turns out that spatial
segregation induces a \emph{drift} in the transitions between
directions, from directions followed by small clusters to directions
of large clusters.  In other words, large clusters are more likely to
gain particles while small clusters tend to loose particles.  This
might seem obvious in a typical coarsening dynamics like the one in the
Ising model, where smaller clusters tend to vanish and the average
size of clusters increases with time.  However, ordering in the
original voter dynamics is quite different because same-state domains
gain and loose particles at the same rate, independent on their size,
and thus coarsening is only driven by fluctuations.  This is due to
the fact that all opinion states are equivalent in the VM on regular
topologies \cite{Suchecki-2005} and, as consequence, the average
fraction of particles in each state is conserved at all times.
Therefore, the drift or net flow of particles between any two states
is zero in the VM.

\begin{figure}[t]
\begin{center}
\vspace{0.5cm}
\begin{tabular}{cc}
\hspace{-1.0cm}
\includegraphics[width=4.0cm, bb=70 -20 550 550]{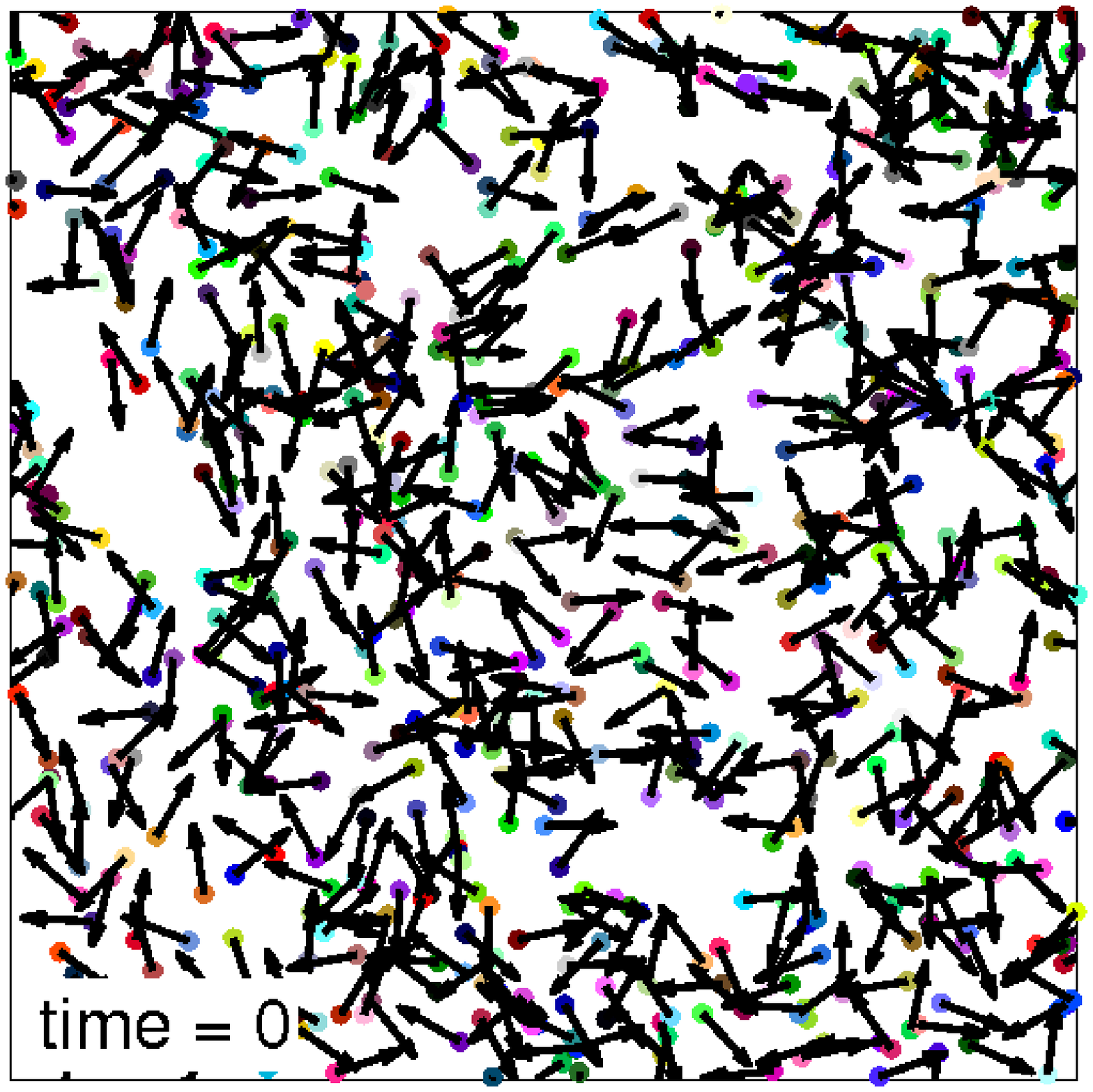}
& \hspace{2.5cm} 
\includegraphics[width=4.0cm, bb=70 -20 550 550]{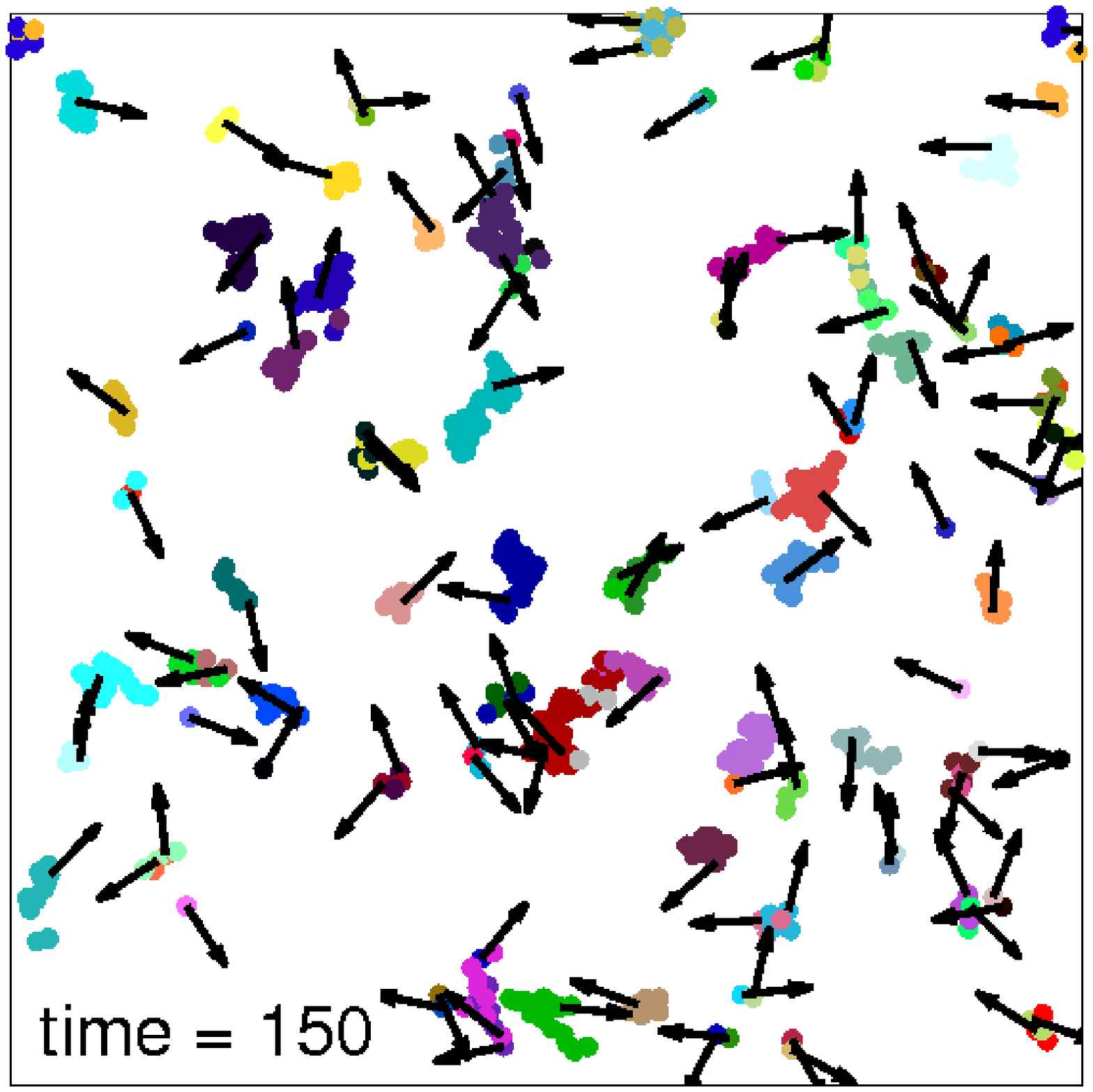} \vspace{0.3cm} \\
\hspace{-1.0cm}
\includegraphics[width=4.0cm, bb=70 0 550 550]{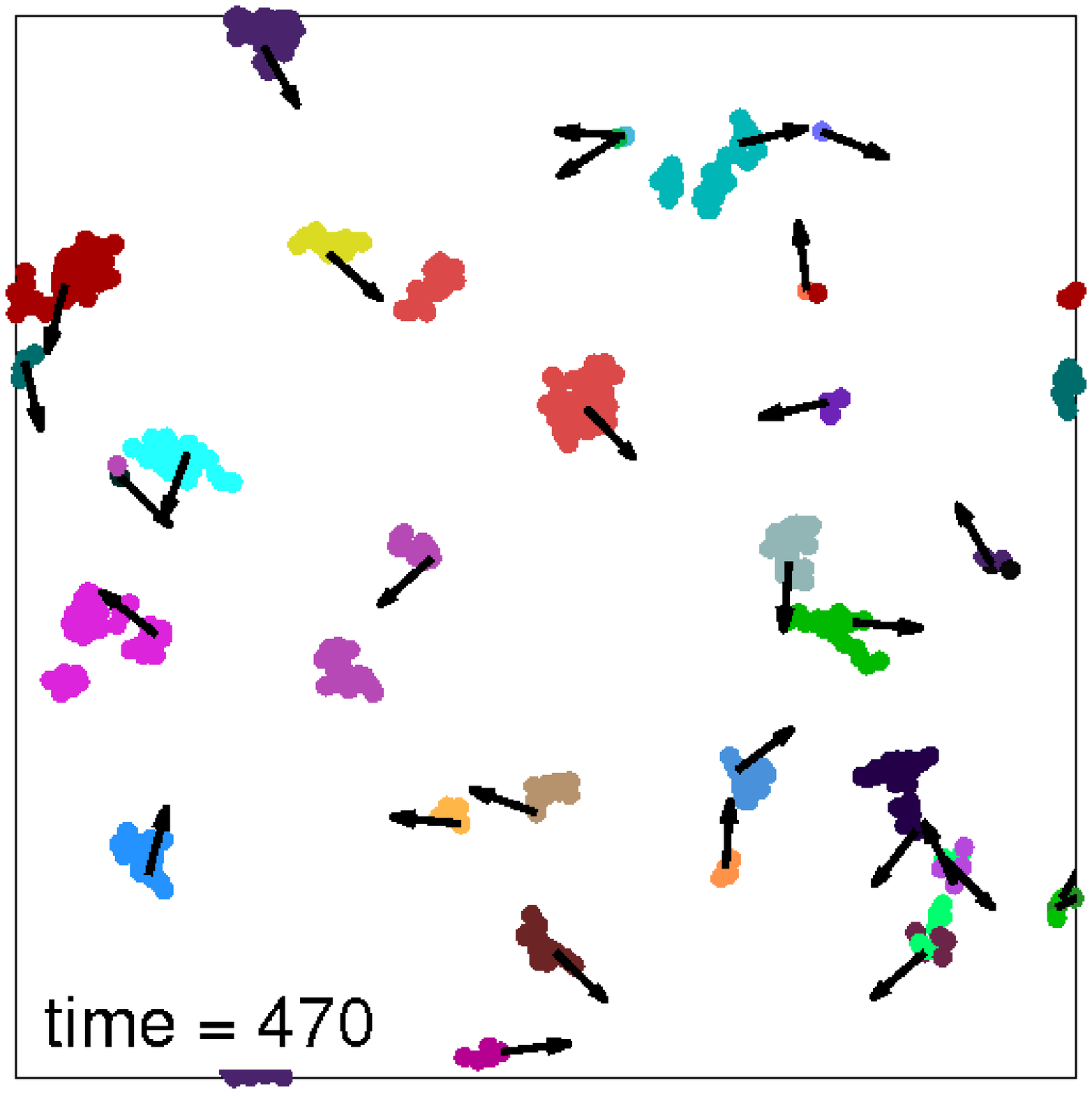}
& \hspace{2.5cm} 
\includegraphics[width=4.0cm, bb=70 0 550 550]{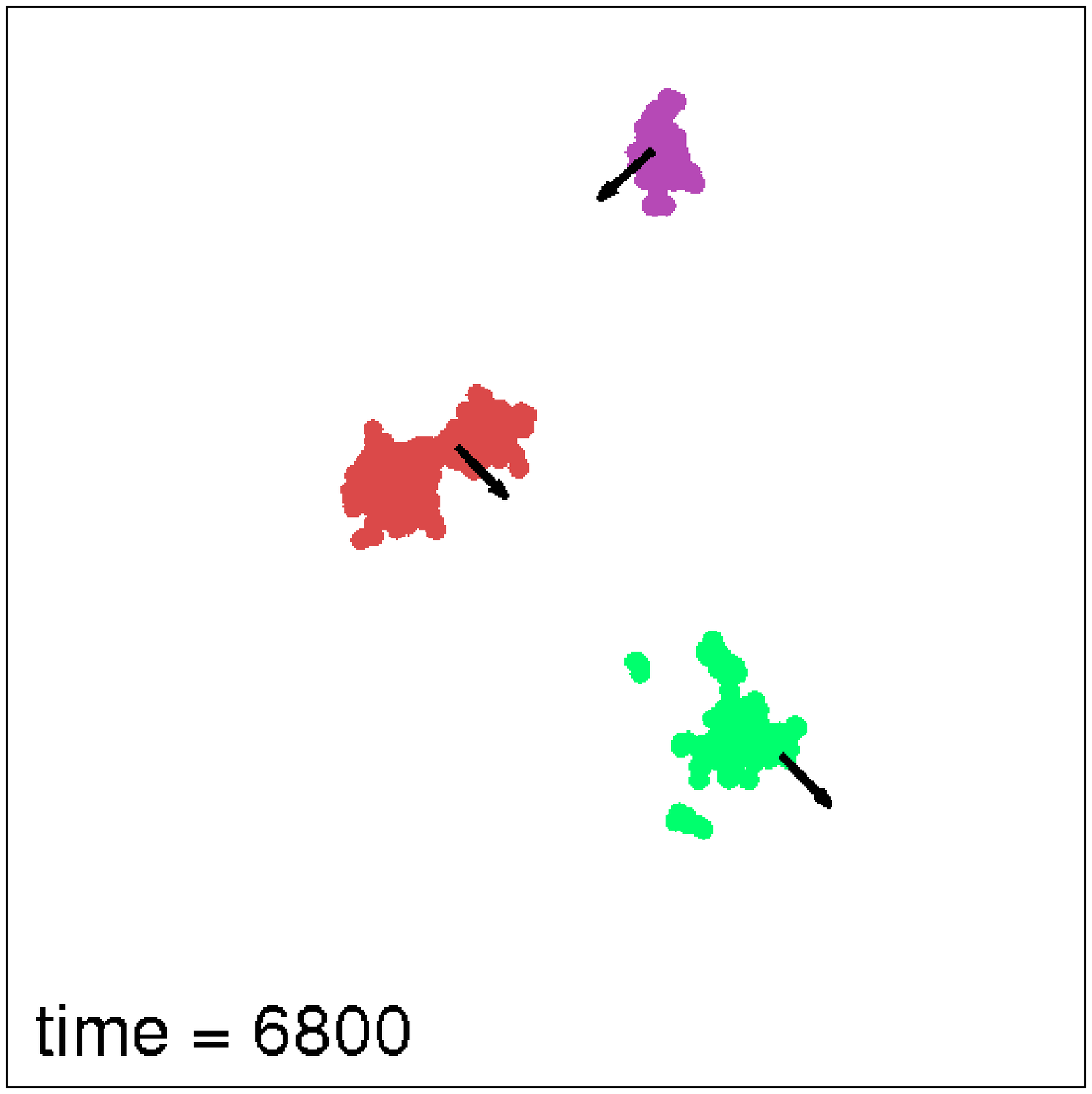} 
\end{tabular}
\caption{Time evolution of four different magnitudes: (a) the net drift $D$ from small to large clusters, (b)
  the time derivative of the mean number of neighbors $\langle k
  \rangle$, (c) the ratio between the mean number of directions in MF ($S_{MF}$)
  and in 2D ($S$) as defined in Fig.~\ref{S-t}, and (d) the covariance
  $\mbox{cov}(m,\langle k \rangle)$ between the size
  of a cluster and its mean degree.  Simulations correspond to systems
  with $N=600$ particles, speed $v=5$ and densities $\rho = 2$
  (circles),  $\rho = 9$ (squares), $\rho = 16$ (diamonds) and $\rho =
  31$ (triangles).}    
\label{drift}
\end{center}
\end{figure}

To study whether the transitions to a given direction $\alpha$ are
correlated with the number of particles with direction $\alpha$ (mass
$m_{\alpha}$), we define the drift from direction $\beta$ to
direction $\alpha$ at time $t$ as
\begin{equation}
D(\beta\to\alpha,t) \equiv \sum_{i=1 / \theta_i = \beta}^N p(\theta_i\to\alpha,t),
\end{equation}
where the sum is over all particles with direction $\theta_i=\beta$,
and $p(\theta_i \to \alpha,t)$ is the  probability that particle $i$
adopts the direction $\alpha$ at time $t$, calculated as the fraction
of $i$'s neighbors with direction $\alpha$.  Then, the net drift from
small to large clusters is defined as 
\begin{equation}
D(t) = \sum_{\alpha} \sum_{\beta} \mbox{sign}(m_{\alpha}-m_{\beta}) \,
D(\beta \to \alpha,t), 
\end{equation}
where $\mbox{sign($x$)}$ is a function that takes the value $1$ ($-1$)
for $x>0$ ($x<0$), and $0$ for $x=0$, thus it assigns a positive
weight to drifts towards larger clusters, and a negative weight to
drifts towards smaller clusters.  Therefore, a positive (negative)
value of $D$ means that, in average, the net drift in the system is
from smaller (larger) to larger (smaller) clusters.  In
Fig.~\ref{drift}(a) we show the time evolution of $D$ averaged over $3
\times 10^4$ realizations, on a system with $N=600$ particles moving
at speed $v=5$ and for various densities.  We observe that $D$ is
larger than zero for all times, showing that there is a net drift from
small to large clusters.  This generates a positive feedback in which
large clusters tend to increase their size while small clusters tend
to shrink, and is in contrast with the MF behavior, where no direction
has a prevalence on the others and thus $D=0$ at all times.
Therefore, the presence of a positive drift breaks the symmetry of the
system and speeds up the evolution towards consensus, as compared to
MF.   We can check that the MF limit is achieved as $\rho$ increases,
where we see that $D$ decreases and is already very small for
$\rho=31$.

As we can see in Fig.~\ref{drift}(b), the shape of the time derivative
of $\langle k \rangle$ agrees quite well with the shape of $D(t)$,
showing a direct relationship between clustering and drift.  Also, $D$
reaches a maximum when $S$ starts to deviate from the MF curve [see
  Fig.~\ref{drift}(c)], indicating that the drift accelerates the
evolution towards consensus.  Figure~\ref{drift} shows only the case
of $v=5$, but we found similar results for $v=0.1$ and $v=20$ as well.
We speculate that the reason why larger clusters tend to increase
their size more rapidly than smaller clusters is because particles
that belong to larger clusters have, in average, a larger number of
neighbors.  And, it is known that in disordered topologies of
interactions like complex networks, the weighted magnetization
($m=\sum_{i=1}^N S_i k_i/N$, with $S_i$ and $k_i$ the state and degree of
node $i$, respectively) is
conserved in the voter dynamics \cite{Suchecki-2005}, rather than the
state magnetization.  This means that
nodes are more likely to copy the opinion state of other nodes with
large degrees, breaking the equivalence between states.

 To check the relation between cluster size and mean degree we
 calculated the time evolution of the covariance $\mbox{cov}(m,\langle
 k \rangle)$ between the  mass $m$ of a given cluster and the mean
 degree $\langle k \rangle$ of the particles that belong to that
 cluster, regardless whether the  neighbors belong to the same or to a
 different cluster.  The evolution of the covariance averaged over
 many realizations is  shown in Fig.~\ref{drift}(d).  We observe a
 positive covariance between $m$ and $\langle k \rangle$ which shows
 that, indeed, larger clusters  have more neighbors, increasing their
 chances to gain more particles.  A similar result was obtained in the
 SVM for which bigger clusters have larger mean degrees as well
 \cite{bagliettoNets}.  The fact that larger clusters tend to have a
 larger mean degree and grow faster, breaks the symmetry of the FVM. 

\section{Discussion and conclusions}
\label{conclusions}

We proposed and studied a flocking model in which self-propelled
particles interact via a simple dynamics of velocity imitation.  We
studied the ordering dynamics of the system and its 
approach to the fully ordered state (consensus) 
$\langle \varphi \rangle=1$.  We found that the dynamics is
characterized by an algebraic increase with time 
($\langle \varphi \rangle \sim t^{1/2}$) during an initial transient,
and a final exponential approach to the ordered state.  Interestingly, a
similar ordering behavior is observed in the SVM model at zero noise,
although the initial algebraic increase happens earlier, and thus the
approach to order is much faster than in the FVM.  We suspect that the scaling
exponent $1/2$ found in the noiseless SVM might be related to the
coarsening exponents found in the ordered phase of the SVM with noise   
\cite{Chate-2008,Dey-2012}.  The ordering in the FVM is related to
the decreasing number of 
different directions of motion in the system, which is well explained
by the MF theory of the multi-state voter model during a first
stage.  On a second stage, and for high enough particle density
$\rho$, the ordering dynamics is faster than in MF due to a break in
the symmetry of directional states that speeds up the dynamics.  As a
consequence, the  
mean time to reach consensus $\tau$ is non-monotonic with $\rho$, that
is, there is an optimal density for which the system reaches
full order in the shortest time.  The shape of the $\tau$ vs $\rho$
curve can be qualitatively explained in terms of three main mechanisms
that act on different density scales.  At low densities, the dynamics
is limited by the sporadic encounters between particles that
introduce a delay in the interactions and lead to large consensus
times.  At intermediate densities, the dynamics undergoes a breaking
in the symmetry of transitions between directions, induced by the
spatial segregation of particles into same-direction clusters.  This
symmetry breaking is enhanced by the motion of particles, generating a
net mass drift from small to large clusters and accelerating the
approach to consensus respect to the MF behavior.  The sublinear
scaling of the mean consensus time with $N$ at the minimum of the $\tau$ vs
$\rho$ curve ($\tau_{\mbox{\tiny min}} \sim N^{0.765}$) is consistent with a
consensus that is faster than in MF.  Finally, at high densities
the MF case of all-to-all interactions is recovered.  

We explored a voter-like dynamics acting on self-propelled agents
subject to {\it metric interactions}, which by definition occur when
two particles are less than a predefined cutoff distance apart.  This
type of interactions was adopted for its simplicity.  Nevertheless, it
is known that some social living organisms, like birds or humans, may
interact in a different way, for instance by considering the first $k$
closest neighbors regardless of the distance to them
\cite{cavagna2008}.  It would be worthwhile to study a system with
voter-like interactions in this context.  In particular, the density
dependence of the dynamics could turn out to be very different with 
respect to the one found in this article, as it happens with the Vicsek
model where the dynamics is essentially independent of the density in
the case of non-metric interactions \cite{Ginelli-2010,Peshkov-2012}.
Finally, a natural extension of the model would consist on the introduction of
noise perturbing the direction of motion of particles as in the SVM.
We plan to study this extended version of the FVM in a future work.

\section*{Acknowledgement}

We acknowledge discussions with Dr. J. Candia, Dr. E. Albano,
Dr. T. Grigera and Dr. E. Loscar.  We acknowledge financial support
from CONICET (PIP 11220150100039CO and PIP 0443/2014) and   Agencia
Nacional de Promoci\'on Cient\'ifica y Tecnol\'ogica (PICT-2015-3628)
(Argentina).  G. Baglietto and F. Vazquez are members of CONICET.

\end{document}